\documentclass[12pt,apj]{emulateapj}

\usepackage{natbib, graphics}

\usepackage{longtable}
\usepackage{amsfonts}
\usepackage{rotating}
\RequirePackage{amssymb}
\RequirePackage{latexsym}
\RequirePackage{graphicx}
\def\feh{\hbox{\rm [Fe/H]~}}

\begin{document}

\title{First Detection of the White-Dwarf Cooling Sequence of the Galactic
Bulge\altaffilmark{1}}

\shorttitle{The white dwarf cooling sequence of the Galactic bulge} 

\shortauthors{Calamida et al.}

\author{A. Calamida$^2$,
K. C. Sahu$^2$, 
J. Anderson$^2$,
S. Casertano$^2$,
S. Cassisi$^3$, 
M. Salaris$^4$,
T.~Brown$^2$,
J. Sokol$^2$,
H. E. Bond$^{2,5}$,
I. Ferraro$^6$,
H. Ferguson$^2$,
M. Livio$^2$,
J. Valenti$^2$,
R.~Buonanno$^3$, 
W.~Clarkson$^7$, and
A. Pietrinferni$^3$}

\altaffiltext{1} {Based on observations made with the NASA/ESA {\it
Hubble Space Telescope}, obtained by the Space Telescope
Science Institute. STScI is operated by the Association of Universities for
Research in Astronomy, Inc., under NASA contract NAS~5-26555.}

\altaffiltext{2}
{Space Telescope Science Institute, 
3700 San Martin Dr.,
Baltimore, MD 21218, USA;
calamida@stsci.edu}

\altaffiltext{3}
{Osservatorio Astronomico di Teramo - INAF}

\altaffiltext{4}
{Astrophysics Research Institute, Liverpool John Moores University}

\altaffiltext{5}
{Department of Astronomy \& Astrophysics, Pennsylvania State University,
University Park, PA 16802, USA; heb11@psu.edu}

\altaffiltext{6}
{Osservatorio Astronomico di Roma - INAF}

\altaffiltext{7}
{University of Michigan-Dearborn}

\begin{abstract}

We present {\it Hubble Space Telescope\/} data of the low-reddening Sagittarius
window in the Galactic bulge.  The  Sagittarius Window Eclipsing Extrasolar
Planet Search field  ($\sim$ 3\arcmin$\times$ 3\arcmin), together with three more
Advanced Camera for Surveys and eight Wide Field Camera 3 fields, were observed in
the $F606W$ and $F814W$ filters, approximately every two weeks for two years, with the
principal aim of detecting a hidden  population of isolated black holes and neutron
stars through astrometric microlensing.  Proper motions were measured with an
accuracy of $\approx$ 0.1 $\rm mas\,yr^{-1}$ ($\approx$ 4 $\rm km\,s^{-1}$) at
$F606W \approx$ 25.5 mag, and better than $\approx$ 0.5 $\rm mas\,yr^{-1}$ 
($\approx$ 20 $\rm km\,s^{-1}$) at $F606W \approx$ 28 mag, in both axes.
Proper-motion measurements allowed us to separate disk and bulge stars and obtain a
clean bulge color-magnitude diagram. We then identified for the first  time a
white dwarf (WD) cooling sequence in the Galactic bulge, together
with a  dozen candidate extreme horizontal branch stars. The comparison between
theory and observations  shows that a substantial fraction of the WDs ($\approx$
30\%) are systematically redder than the  cooling tracks for CO-core H-rich and
He-rich envelope WDs. This evidence would suggest the  presence of a significant
number of low-mass WDs and WD - main sequence binaries in the bulge.  This
hypothesis is further supported by the finding of two dwarf novae in
outburst, two short-period ($P \lesssim 1$~d) ellipsoidal variables, and a few
candidate cataclysmic variables  in the same field. 

\end{abstract}


\maketitle

\section{Introduction}\label{intro}

Most stars end their lives as white dwarfs (WDs). The characterization of WD
populations is a valuable tool for understanding the formation history of 
different components of our Galaxy. Our knowledge is most extensive for the WD
population of the Galactic disk, in which numerous WDs have been discovered
through imaging surveys and characterized through spectroscopy
\citep{eisenstein06, kepler07,koester09}.  An updated catalog from the Sloan
Digital Sky Survey Data Release~7, including 12,843 DA (H-rich envelope) and 923
DB (He-rich envelope) WDs, has been presented recently by \citet{kleinman13}. 
They find a mean mass of $\sim$0.6$M_{\odot}$ for DA and $\sim$0.68 $M_{\odot}$
for DB WDs, consistent with CO cores.  There is also a secondary peak in the
Galactic-disk WD mass distribution around $0.4 \,M_{\odot}$
\citep{kepler07,rebassa11}.  These low-mass WDs likely have helium cores and,
given that in standard stellar evolution they cannot be formed from single stars
in less than a Hubble time, they have been considered to result from
close-binary interactions after a common-envelope phase (post common-envelope binaries, PCEBs), 
or of a merger of two very low-mass He-core WDs following a second CE ejection \citep{han08}. The
binary scenario is supported by the finding of numerous low-mass WDs in binary
systems in the field, with the companion being another WD, a neutron star, or a
subdwarf B star \citep{marsh95,maxted02}.  However, single He-core WDs have also
been observed \citep{marsh95, kilic07}, raising the possibility that severe mass
loss on the first ascent of the giant branch can prevent ignition of helium
burning. Extremely low-mass He-core WDs ($M \simeq 0.2\, M_{\odot}$) have only
been found in binary systems \citep{kilic12}.

In Galactic globular clusters (GGCs) a substantial population of He-core WDs has
been observed in $\omega$ Cen \citep{monelli05, calamida08, bellini13}, NGC~6752
\citep{ferraro03}, and NGC~6397 \citep{strickler09}, as well as in the old
metal-rich open cluster NGC~6791 \citep{kalirai07, bedin08}. He-core WDs in
clusters show systematically redder colors than CO-core WDs. 

In the Galactic bulge, the bulk of the population is as old as GGCs but as
metal-rich as the Galactic disk \citep[hereafter CL08]{zoccali03, clarkson08},
with a stellar space density closer to that in the disk than in GGCs.  If  age plays a
major role in the formation of He-core WDs, we would expect an enhanced
frequency of such objects in the Galactic-bulge population. If stellar density
and/or metallicity are the determinants instead,  the bulge population of
He-core WDs, as well as their binary fraction, should be consistent with those
in the Galactic disk. 

In this manuscript we present the first observational detection of the WD cooling
sequence in the Galactic bulge, and discuss the implications for the origin of
low-mass WDs. We have also discovered several cataclysmic variable (CV)
candidates including two dwarf novae, as well as two short-period
ellipsoidal binary systems.

\section{Observations and photometry}\label{obs}

We observed the Sagittarius Window Eclipsing Extrasolar Planet Search (SWEEPS) field
($l = 0^\circ, b = -2\fdg65$) in the Galactic bulge in 2004 and again in 2011, 2012 and 2013
with the {\it Hubble Space Telescope\/} ({\it HST}), using the Wide-Field Channel of the 
Advanced Camera for Survey (ACS, proposals GO-9750, GO-12586, PI: Sahu).
The SWEEPS field covers $\approx 3\farcm3 \times 3\farcm3$ in a region of relatively low extinction in
the bulge ($E(B-V) \lesssim 0.6$~mag; \citealt{oosterhoff}).
The 2004 observations were taken in the $F606W$ (wide $V$) and $F814W$ (wide $I$) filters over the
course of one week (for more details see \citealt{sahu06}). The new data were
collected between October 2011 and October 2013, 
with a $\sim$ 2-week cadence, for a total of 60 $F606W$- and 61 $F814W$-band images.  
The 2011--2012--2013 (hereafter 2011--13) dataset was reduced using a software 
program that performs simultaneous point-spread
function (PSF) photometry on all the images (Anderson et al., in preparation). We
adopted the 2004 photometric zero-points to calibrate the data to the Vegamag
system, producing a catalog  of $\approx$ 340,000 stars down to $F606W \approx 31$
mag and the deepest color-magnitude diagram (CMD) so far published in the
direction of the Galactic bulge. 

A dozen images of the SWEEPS field  were also collected in the $F625W$ and $F658N$ filters
with the ACS camera during March 2011 (proposal GO-12020, PI: Clarkson). 
The reduction of this dataset was performed by following the same procedure 
described above. The final catalog was calibrated to the Vegamag
system and includes $\approx$ 200,000 stars down 
to $F625W \approx 27$ mag.

Our reduction software provides quality parameters such as the dispersion of the 
individual photometric and astrometric observations, the similarity of the object to the shape 
of the PSF and the degree of contamination of the object by neighboring stars;  
we use these quality parameters to cull the photometry to keep only well-measured objects. 
The left panel of Fig.~1 shows the $F606W,\, F606W
- F814W$ CMD for all the stars with high-quality photometry.
The CMD shows a few interesting features:  i) there is a group of stars
systematically redder than the bulge main-sequence (MS) that belong to the
(closer) disk population; ii) there are about a dozen candidate EHB stars clearly
visible in the CMD at 20 $\lesssim F606W \lesssim$ 22 mag and $F606W - F814W
\approx$ 0.3 mag; iii) there is also an indication of a WD cooling
sequence starting at $F606W \approx$ 22.5 mag and extending below, in the color
range 0 $\lesssim F606W - F814W \lesssim$ 1.5 mag. 

EHB stars have been identified in Baade's window of the Galactic bulge by \citet{zoccali03}, 
and spectroscopically characterized as hot subdwarf stars by \citet{busso05}.
A WD population has never been identified, so it is important to 
determine whether they are indeed WDs and whether they belong to the bulge.

\begin{figure*}
\begin{center}
\label{fig1}
\includegraphics[height=0.7\textheight,width=0.57\textwidth,angle=90]{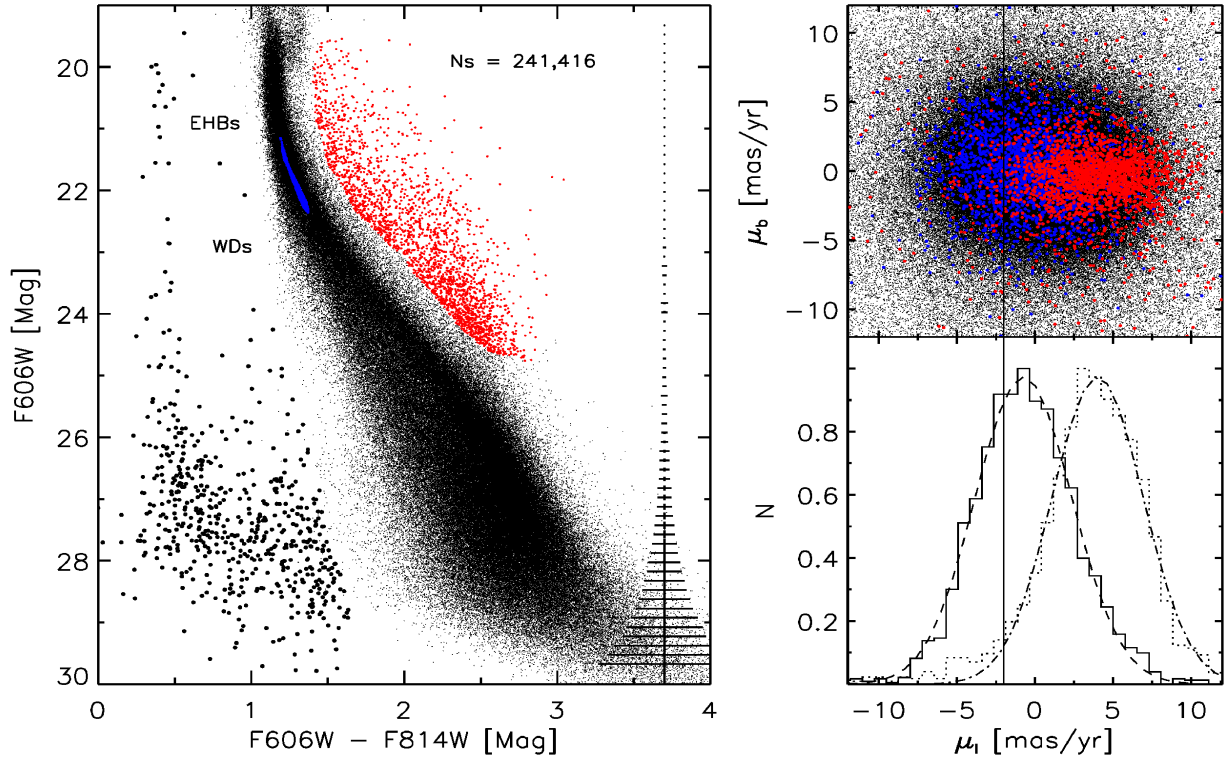} 
\caption{Left -- $F606W,\ F606W - F814W$ CMD of stars selected to have high-quality photometry in the SWEEPS field. 
Candidate EHB and WD stars are marked with larger dots. The color cut off was chosen not to 
exclude candidate WD-MS binaries.
Samples of candidate bulge and disk stars are marked with blue and red dots, respectively. 
Right -- Top: PM diagram of the selected stars, with samples of candidate bulge and disk stars marked.
Bottom: normalized PM histograms of Galactic longitude components for candidate
bulge (left) and disk (right) stars. The two Gaussians fitting the distributions are over-plotted.}
\end{center}
\end{figure*}

\subsection{A clean white dwarf bulge sample}\label{clean}

Stars could be in the WD region of the CMD because they are bona fine bulge or disk WDs, or 
because they are closer and less reddened disk MS stars. 
It is important then to consider a bulge sample devoid of disk stars, since some disk stars with
much less reddening may appear bluer, and hence masquerade as WDs.  
To obtain a clean bulge sample, we estimated the proper motions (PMs) of the stars using the 2004 and 2011--13 datasets.
By comparing the positions of stars in the two epochs we estimated the PMs for $\approx$
200,000 stars down to $F606W \approx$ 28 mag,  with an accuracy $\lesssim$ 0.5 $\rm
mas\,yr^{-1}$ ($\lesssim$ 20 $\rm km\,s^{-1}$) in both axes. PMs are then  projected along
the Galactic coordinates shown in Fig.~1 (top right panel).  

We then used the CMD to select two samples of candidate stellar populations belonging to
the bulge (blue dots) and the  disk (red), following a procedure similar to that adopted by CL08.
The two samples were selected in different magnitude and color ranges to keep the sample 
sizes similar, and keep the contamination of the bulge sample by disk stars minimal.
The PM histograms of the Galactic-longitude components for these two populations are
shown in Fig.~1 (bottom right). The two populations are clearly separated, with the bulge
peaking at $\mu_l \approx$ 0 $\rm mas\,yr^{-1}$, and the disk peaking at $\mu_l \approx$ 
4 $\rm mas\,yr^{-1}$.  The PM distribution of disk and bulge stars is a consequence of
Galactic rotation, a detailed discussion of which is given in CL08.  We adopted a cut at
$\mu_l \le -2 \,\rm mas\,yr^{-1}$ to select a nearly pure bulge sample: this selection allows us to keep
$\approx$ 30\% of bulge members while the residual contamination of the sample by disk stars
is $\lesssim$ 1\%. Fig.~2 shows the CMD of the resulting pure-bulge sample where the
photometry of stars brighter than $F606W = 19.8$ mag (which is the saturation limit for
the 2011--13 dataset) is taken from the 2004 dataset, while the photometry of the fainter
stars is from 2011--13. 
About 70\% of the bulge WDs were indeed rejected because of the PM 
selection, although most of the WDs in the selected magnitude and color range belong to the bulge.
To further select candidate WDs, we individually checked each star with ROMAFOT
\citep{buonannoiannicola89} on the $F606W$ and the $F814W$ median images.  
Some of the candidate bulge WDs turned out to be too close to or to lie on the spikes 
of saturated stars, and a few others were too faint to be reliably measured ($F606W >$ 28 mag); these were rejected from the sample. 
We note that the PM selection has led to a reduction of the number of brighter 
 (22 $< F606W <$24 mag)  WDs from 8 to 4 objects, 2 of which happen to be CVs.
Table~1 lists coordinates, magnitudes, and PMs for the 72 well-measured bulge WDs. 
Only these well-measured WDs are included in Fig.~2, which clearly shows a sequence 
of objects ranging from 22.5 $\lesssim F606W \lesssim$ 29 mag, with most of them 
having a color $F606W - F814W \lesssim$1.0 mag.
The only known class of objects that can occupy this part of the CMD are bulge WDs
and unreddened disk stars. Since almost all ($\gtrsim$ 99\%) disk stars were removed by 
the PM selection (see \S 2.1) the only remaining possibility is bulge WDs.
Furthermore, faint blue galaxies, which play the role of serious contaminants in the WD sequence observed
in GGCs \citep{richer08}, are absent in our sample 
since the large reddening by the Galactic disk in this direction
serves as a natural filter to remove any background galaxies.

Theoretically, we expect to observe a WD cooling sequence in the bulge,
since the bulge is old and formed over a relative short timescale, as suggested by
the scarcity of younger stars (CL08). 
It is then natural to ask: why was this bulge WD sequence not identified until now?  
The reason is simple. It was essential to separate the disk stars in order to identify them as
bulge WDs. Since the WDs are intrinsically faint, the PMs of this sample was not known. 
Our deep HST observations taken at 2 epochs were essential to separate the faint disk stars from
the bulge WDs.

\section{Discussion}\label{discussion_conclu}
\subsection{Theoretical models}
In order to characterize the bulge stellar population, we compared the
PM-cleaned bulge CMD with evolutionary predictions.
We used the BaSTI\footnote{http://albione.oa-teramo.inaf.it/} \citep{pietrinferni04,pietrinferni06}
stellar-evolution database to fit isochrones to the CMD. 
Evolutionary predictions were transformed  to the observational plane by adopting 
the color--$T_{\rm eff}$ relations provided by \citet{hauschildt99} for $T_{\rm eff}  \le 10,000$~K, 
while at larger $T_{\rm eff}$ we adopted the relations published by \citet{bedin05a}. 
A distance modulus of $DM_0 = 14.45$ mag and a mean reddening of $E(B-V) = 0.5$ 
were adopted  \citep{sahu06}. Extinction coefficients were estimated by applying the \citet{cardelli89}
reddening relations and by adopting a standard reddening law, $R_V = A_V/E(B-V) = 3.1$, finding
$A_{F606W} = 0.922 \,A_V$,  $A_{F814W} = 0.55 \,A_V$, and  $E(F606W - F814W)= 1.14\, E(B-V)$. 

Solid lines in Fig.~2 show isochrones for an age of $t = 11$~Gyr and a scaled-solar mixture with 
different chemical compositions. The blue line corresponds to $Z = 0.03, Y = 0.288$ ($\feh$ = 0.26), the red
to $Z = 0.0198, Y = 0.273$ ($\feh$ = 0.06), and the green to $Z = 0.008,  Y = 0.256$ ($\feh = -0.35$).
Zero-age horizontal branches (ZAHBs) are plotted for the most metal-poor (green) and the most metal-rich 
chemical compositions (blue). We adopted these mixtures based on the medium-resolution spectroscopy 
of 93 turn-off (TO), red-giant (RG) and MS stars in the SWEEPS field, collected with FLAMES/VLT (ESO, cyan dots).  
The distribution spans a range of $-1.0 < \rm[M/H] < 0.8$, and shows three main peaks at 
$\rm[M/H] \approx -0.5$,  0.0 and  $\approx$0.25. This distribution is in fairly good agreement with 
the spectroscopic metallicity distributions found by \citet{hill11}, \citet{bensby13}, and \citet{ness13} for the bulge.

To fit the bulge WD cooling sequence we adopted the BaSTI cooling tracks for DA and DB CO-core WDs and 
the models of \citet{althaus09} for He-core WDs.
For $t \approx 11$~Gyr, and solar metallicity, the mass of the stars at the TO  is $\approx 0.95 \,M_{\odot}$ 
and the theoretical initial-to-final mass relationships predicts WD masses of $\sim$0.53--$0.55 M_{\odot}$ \citep{weissferguson09}.  
This prediction is observationally supported by the spectroscopic measurements of bright WDs in the GGC
M4 by \citet{kalirai09}, for which they find a mean mass of $\sim$$0.53\,M_{\odot}$.  
So we used cooling tracks for DA and DB CO-core WDs with mass $M = 0.54 \,M_{\odot}$ \citep{salaris10}, 
and He-core WD tracks for a mass of $0.23 \,M_{\odot}$. 
We applied to the cooling tracks the same distance modulus and reddening adopted for the isochrones.

\begin{figure*}
\begin{center}
\label{fig2}
\includegraphics[height=0.7\textheight,width=0.57\textwidth,angle=90]{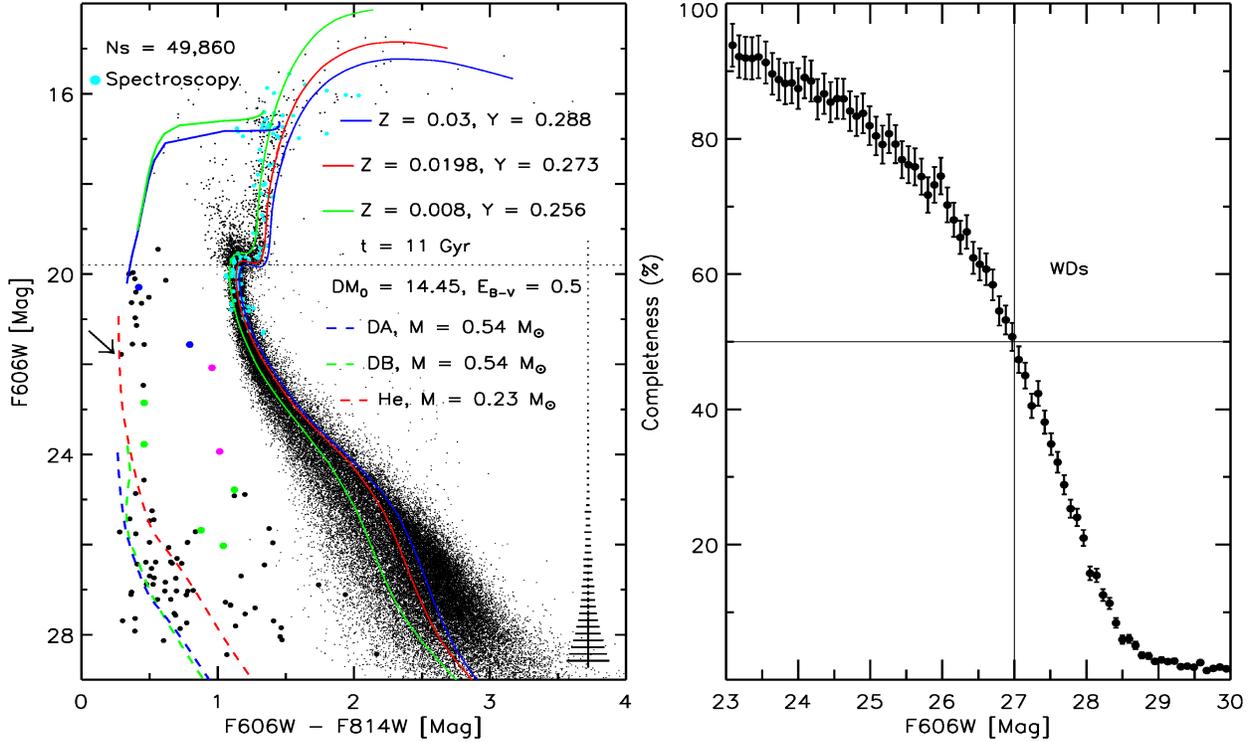} 
\caption{Left: PM-cleaned bulge CMD.
Solid lines display cluster isochrones for the same age but different chemical composition. 
ZAHBs and cooling sequences for CO- and He-core WDs are plotted. 
Cyan dots mark stars with spectroscopic observations.
The arrow marks the reddening vector direction. Ellipsoidal variables and dwarf novae
are marked with blue and magenta dots, respectively. Green dots mark CV candidates.
Right: completeness of bulge WDs. Note that 70\% of the bulge WDs were rejected because of the PM 
selection, although most of the WDs in the selected magnitude and color range belong to the bulge.
Error bars are also labelled.
}
\end{center}
\end{figure*}

The comparison between theory and observation shows that old
scaled-solar isochrones spanning almost 1 dex in metallicity, $-0.4 \lesssim
\feh \lesssim 0.3$, fit the bulge MS and RGB quite well over most of the color range.
However, there are stars systematically bluer or redder than the most
metal-poor and metal-rich isochrones. 
Part of this residual color spread is due to photometric 
errors, differential reddening and depth effects. 

Fig.~2 also shows that cooling tracks for DA (dashed blue line) and DB (dashed green) CO-core 
WDs, which are expected to be shifted up to $\pm$0.5 mag due to the bulge depth, are unable to
reproduce the entire color range of the observed WD cooling sequence. An
increase in the mean mass of the WDs would move the models towards bluer
colors, further increasing the discrepancy. We then  assume the presence of a
fraction of low-mass, $M \lesssim 0.45\,M_{\odot}$, WDs in the bulge.  The lower
mass He-core cooling track for $M = 0.23 \,M_{\odot}$ (dashed red) fits the red
side of the bulge WD sequence,  but it is not able to account for the reddest WDs. 
Note that empirical evidence shows that the lower mass 
limit for WDs is $\approx 0.2\,M_{\odot}$ \citep{kepler07}. 

\subsection{The color spread of the WD cooling sequence}
To properly characterize the color spread and the completeness of the bulge
WD cooling sequence, shown in the right panel of Fig.~2, we performed several
artificial star (AS) tests to estimate the magnitude and color dispersion of the
sequence due to photometric errors and to the reduction and selection techniques adopted. 
We randomly added $\approx$160,000 artificial WDs to all images, with magnitudes
and colors estimated by adopting a DA cooling track with $M = 0.54 \,M_{\odot}$,
and by adding a distance modulus of  14.45 mag and a mean reddening of
$E(B-V) = 0.5$. Artificial stars are added and recovered on the images one at a
time, not to affect the crowding, by using the same reduction procedures adopted
earlier.  We then estimated the magnitude and color spread of the artificial WD
cooling sequence. This information was used as input to produce a synthetic
sequence from the DA cooling track. After correcting the model absolute
magnitudes for distance modulus and reddening, we have drawn randomly 30,000 ages
with a uniform probability distribution, within the age range corresponding to
the observed WD magnitude range. The resulting magnitudes were then perturbed
with a Gaussian photometric error obtained from the AS tests.  

Fig.~3 shows the synthetic WD sequence, which also includes the effect of differential reddening.
We added to the star magnitudes extinction values simulated as clumps with a
maximum peak of $A_V \sim$ 0.6 mag on top of a uniform reddening of $A_V \sim
1.5$~mag.  The assumption of an extinction increase up to $A_V = 1$~mag for
$\approx$ 30\% of the WDs, which would explain the presence of the reddest WDs,
must be discarded since we do not see any evidence for the presence of such 
differential reddening along the MS for a similar fraction of stars. 

Candidate bulge WDs are over-plotted as red dots on the synthetic cooling
sequence of Fig.~3,  together with a DA CO-core (blue line) and a He-core (red)
track. Fig.~3 shows that differential reddening added to photometric errors
cannot account for the entire color spread of the observed WD sequence in the
bulge, by assuming only a CO-core WD population with a mean mass of
$\sim$$0.54\,M_{\odot}$. It is worth noting that if the spread were due
only to the aforementioned factors, we would observe an equivalent number of
WDs bluer than the cooling tracks. 

The agreement between theory and observations improves by assuming that a fraction 
of the WDs have lower masses. This moves the theoretical tracks bright-ward and to the red, 
thus improving the agreement with our observations. These stars could be He-core WDs
as described in \S 3.1, but a fraction of them could also be low-mass CO-core WDs. 
The recent theoretical calculations of \citet{pradamoroni09} showed indeed that CO-core WDs with
masses  down to $\sim$$0.33\,M_{\odot}$ can form in high-density environments,
when a strong episode of mass-loss  occurs along the RGB, due to binary
interactions. However, this theoretical scenario requires very fine-tuned initial conditions,
suggesting that these stars are more likely to be He-core WDs.

\subsection{Binaries as precursors of He-core WDs}
It is not possible to explain the presence of He-core WDs, which are expected to
have masses $\lesssim 0.45 M_\odot$, as a result of single-star evolution in less than a Hubble time. 
As described earlier, a natural explanation for the presence of such 
WDs is that they result from close-binary interactions after a common-envelope phase, 
or from a merger of two very low-mass He-core WDs following a second CE ejection.
In this scenario, a small number of the He-core WDs are expected
to show signatures of variability consistent with WDMS and CV binaries, 
or ellipsoidal variables. 

To look for the presence of binaries in the bulge WD sample, 
we checked the brightest WDs ($F606W \lesssim 26$~mag)  for variability, 
and we indeed identified two ellipsoidal variables (blue dots in Fig.~2), two dwarf novae in 
outburst  (magenta dots in Fig.~2 and 3), and five candidate CVs in quiescence 
(green dots in Fig.~2 and 3). These findings support the presence of He-core 
WDs in the bulge. Moreover, the luminosity of the accretion disk and the companion in the WDMS 
systems moves the WDs towards redder colors in the optical CMD \citep{darnley12}, 
which would be consistent with our observations of the very red WDs.

The $V$ and $I$ light curves of one ellipsoidal variable based on the 2004 dataset are shown
in Fig.~4. The estimated period and amplitude are $P = 2.258\times10^{-1}\pm2\times10^{-5}$
days  and $A = 5.1\times10^{-2}\pm9\times10^{-4}$ mag. By assuming that the primary star is a
WD and the companion a MS star, we estimated a  semi-major axis $a = 0.007$ AU, and masses of
$\approx$ 0.3--$0.4 \,M_{\odot}$ and $\approx$ 0.6--$0.7\,M_{\odot}$ for the two components,
respectively.   An accurate estimate of the masses of the components of these systems  needs
radial velocities; we are in the process of collecting such data through spectroscopic
observations with GMOS at GEMINI South.  It is noteworthy that the $V-I$ color of the variable
is constant  (see Fig.~4), which means that we cannot be observing a reflection effect, i.e.
the heated hemisphere of the MS companion as in the case of the WDMS binary  HS 1857+5144
\citep{aung07}. This ellipsoidal variable is relatively bright with $F606W \sim 21.6$ mag;
stars belonging to the reddest WD cooling sequence could be the fainter counterparts of this
object. Unfortunately, we do not have 2004 time-series photometry for the fainter candidate
CVs to verify this hypothesis.  The 2011--13 dataset has a two-week observation cadence, so we
were unable  to probe all possible periods for our candidate variables.
candidates for short-period variability.  Population-synthesis models predict that most  WDMS
systems have periods in the range 2--30 h \citep{yungelson94}. Moreover, models from
\citet{iben97} predict that $\approx$ 75\% of the WDs in close binaries are low-mass He-core
WDs.  This scenario is observationally supported by \citet{rebassa11} who measured masses for
$\sim$ 200 wide WDMS systems and PCEBs; they find that the mass distribution of the complete
sample of binaries is bimodal, with a main  peak at $M \sim$ 0.55 $M_{\odot}$ and a secondary
one at $M \sim$ 0.40 $M_{\odot}$,  while the distribution of the PCEBs shows a concentration
of systems towards  the low-mass end, with only few binaries with $M \gtrsim$ 0.55 $M_{\odot}$
(see their Fig.~1).

\begin{figure}
\begin{center}
\label{fig3}
\includegraphics[height=0.38\textheight,width=0.5\textwidth]{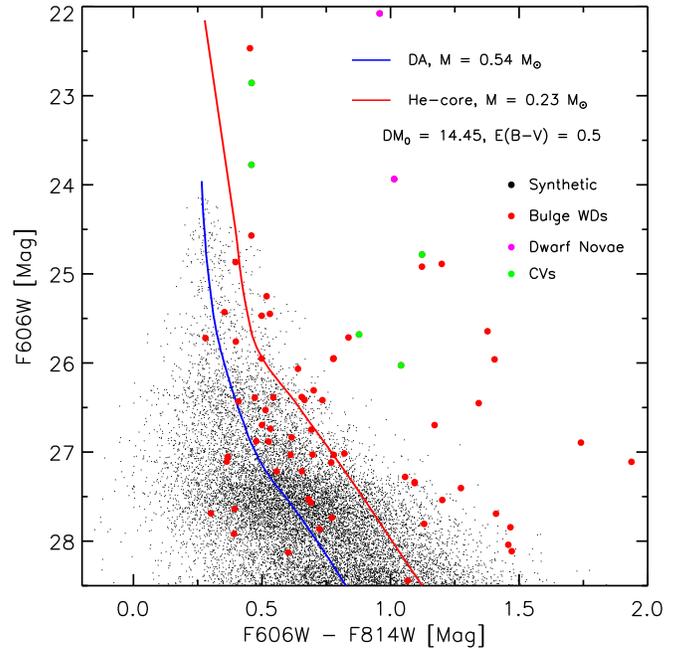}  
\caption{$F606W,\ F606W - F814W$ CMD of $\sim$ 30,000 synthetic WDs. 
Bulge WDs are over-plotted with red dots. Green and magenta dots
mark candidate CVs and the dwarf novae.
CO-core DA (blue line) and an He-core (red) cooling tracks are over-plotted.}
\end{center}
\end{figure}

\begin{figure}
\begin{center}
\label{fig4}
\includegraphics[height=0.35\textheight,width=0.5\textwidth]{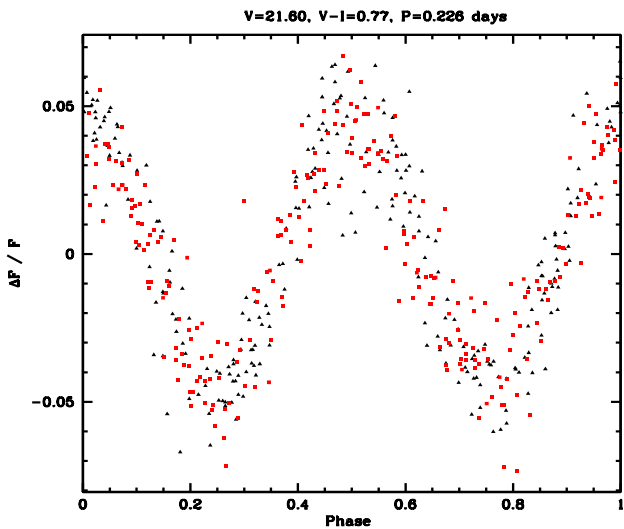} 
\caption{$F606W$ ($V$, red rectangles) and $F814W$ ($I$, black triangles) phased light curves of one of the ellipsoidal variables identified in the SWEEPS field. This variable is located
between the bulge EHB-WD sequences and the MS (see Fig.~1).}
\end{center}
\end{figure}

WDMS binaries and PCEBs might exhibit $H_{\alpha}$-excess \citep{rebassa07}.  To check for
the presence of $H_{\alpha}$-excess among our sample of bulge WDs, we matched the $F606W,
F814W$-band 2011--2013 dataset with the $F625W, F658$-band catalog, finding $\approx$
200,000 stars in common.   Using the PMs from the 2011--2013 dataset, the matched
catalog was cleaned to keep only the bulge stars.  

Fig.~5 shows the $F625W, F658N - F625W$ PM-cleaned CMD for the SWEEPS field. The $F625W,
F658N - F625W$ CMD is designed to specifically highlight the  presence of $H_{\alpha}$
bright stars. In this plane EHBs and the WDs  are expected to be towards the right/red of
the MS since they have stronger $H_{\alpha}$ absorption line. 48 out of 72 bulge WDs and
10 out of a dozen EHBs were identified in this sample and  are marked as larger filled dots
in the figure.  The two ellipsoidal variables (blue dots), the dwarf novae (magenta) and 4
out of 5 candidate CVs (green) were identified too. WDs fainter than $F606W \le$ 27 mag were
not detected either in the $F625W$- or in the $F658N$-band. The same WD cooling tracks
adopted to fit the other CMDs are over-plotted.  The figure shows that the WD cooling
tracks agree with the observations within the uncertainties in  the magnitude range 22
$\lesssim F625W \lesssim$ 27 mag. However, there is a  fraction of WDs that is
systematically towards the left/blue of the CO-core and He-core  cooling tracks.  The
photometric error is $\sigma_{F606W - F814W} \lesssim$ 0.2 mag at $F625W \approx$ 25.5 mag
(shown by a dotted horizontal line in the figure), but some WDs are  shifted towards the
blue of the CMD by more than 2 $\sigma$. 

For magnitudes brighter than $F625W =$ 25.5, there are a total of 17 WDs.  Six of these WDs,
including both the observed dwarf novae,  show $H_{\alpha}$ excess, with $F658N - F625W
\lesssim$ -0.3 mag.  This implies that $\approx$ 30\% of the WDs show $H_{\alpha}$ excess. 
This result further supports the hypothesis of the presence of a fraction of PCEBs and WDMS
binaries  in the bulge.  For magnitudes fainter than $F625W =$ 25.5, there is a trend of the
WD cooling sequence towards bluer colors. This may be either due to blending with other
stars, presence of stars with $H_{\alpha}$ excess as described above, or other yet unknown
effects. It is worth noting that one of the CV candidates ($F606W \sim$ 24.5 mag) shows very
strong $H_{\alpha}$ absorption, with $F658N - F625W \approx$ 0.8 mag. This object resembles
the three CVs  identified by \citep{taylor01} in the NGC~6397, which all have $F658N -
F625W >$ 0.5 mag.

\begin{figure}
\begin{center}
\label{fig5}
\includegraphics[height=0.38\textheight,width=0.5\textwidth]{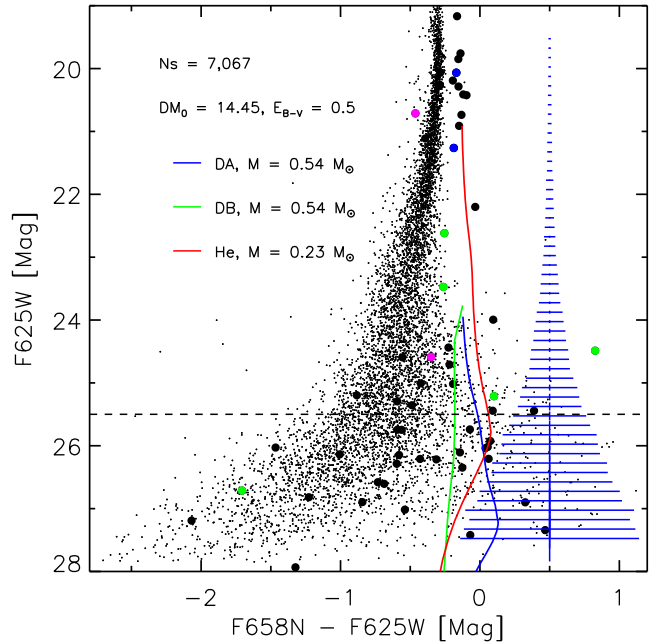}
\caption{$F625W,\ F658N - F625W$ CMD for the SWEEPS field.
Bulge WDs and EHBs are the larger black dots. 
Green and magenta dots mark candidate CVs and the dwarf novae, while
the blue dots mark the two ellipsoidal variables.
CO-core DA (blue line), DB (green) and He-core (red) cooling tracks are over-plotted.
Error bars are also labelled.
}
\end{center}
\end{figure}

\subsection{Number counts}
To further investigate the potential presence of a substantial fraction of He-core WDs
in the bulge, we compared WD counts for $F606W \le 27$~mag, corrected for
completeness, to MS counts across the TO region ($19.9 < F606W < 20.15$~mag), where
the evolution of the star is faster and almost independent
of the initial mass function \citep{calamida08}.

Note that we are taking into account only 30\% of bulge stars due to our PM
selection,  but this does not matter for a comparative analysis since the same PM
selection should affect equally all evolutionary phases.

We selected the brightest PM-selected WDs ($22.5 < F606W < 27$~mag) in seven 
magnitude limits corresponding to $F606W <$ 24, 24.5, 25, 25.5, 26, 26.5, and 27, and counted 
how many stars there are up to each limit.
We then associated to each magnitude limit a WD cooling time, $t_{WD}$, from the 
BASTI DA CO-core cooling track for a mass of $0.54\,M_{\odot}$.
It is interesting to note that the cooling time at $F606W \approx$ 28 mag 
for this sample of WDs is $\approx$ 500 Myr, assuming most stars are 
$0.54\,M_{\odot}$ DA CO-core WDs, and assuming a distance modulus and reddening 
of $DM_0 = 14.45$ mag and $E(B-V) = 0.5$ mag, respectively.

The mean MS mass at the TO magnitude range $19.9 < F606W < 20.15$~mag  is
$0.95\,M_{\odot}$, according to a BASTI scaled-solar isochrone for  $t = 11$ Gyr and
$Z = 0.02$, and the same distance modulus and reddening as described above. 
We then adopted a BASTI evolutionary track for the same composition and for
a mass of  $0.95\,M_{\odot}$ to estimate the time that takes for MS stars to 
cross the magnitude range from $F606W = $19.9 to 20.15 as $t_{MS} \sim 1 \times 10^9$ years.
     
The comparison between observed star counts corrected for completeness, 
$N_{WD}/N_{MS}$, and the theoretical lifetime ratios, $t_{WD}/t_{MS}$, shows that  we
are observing about a factor 2 more WDs than predicted if all the WDs are CO-core.

We now assume the presence of a fraction of $\sim$ 30\% He-core WDs with a mass
of  $0.4\,M_{\odot}$ in the sample based on the discussion above, for which the
cooling times  are significantly longer compared to those of CO-core WDs.  We then
estimate the total WD cooling time as: $$t_{tot} = 0.30 \times t_{cool}(He-core) +
0.70 \times t_{cool} (CO-core)$$ The observed and theoretical ratios then agree within
the uncertainties  taking into account a 10\% uncertainty in the WD cooling times and
MS lifetimes, and the Poisson statistics on the number counts.

This argument thus further strengthens the hypothesis of the presence of a fraction of
He-core WDs in the Galactic bulge.

\begin{deluxetable*}{lcccccc}
\tablecaption{List of the candidate bulge WDs in the SWEEPS field. The first two stars are the ellipsoidal variables.$\rm^a$}           
\tablehead{
\colhead{ID} & \colhead{RA (J2000)} &
\colhead{Dec (J2000)} & \colhead{F814W} & \colhead{F606W} &
\colhead{$\mu_b$} & \colhead{$\mu_l$} \\
\colhead{  } & \colhead{hours} & \colhead{degrees} &
\colhead{[mag]} & \colhead{[mag]} & \colhead{[mas $\rm yr^{-1}$]} & 
\colhead{[mas $\rm yr^{-1}$]} 
}
\startdata
   1   & 17:59:05.17   & -29:11:36.99   &  19.87      &  20.29     &      0.38     &    -0.87   \\
   2   & 17:59:08.06   & -29:12:55.37   &  20.77      &  21.56     &     -0.47     &    -7.23   \\
   3   & 17:59:02.45   & -29:10:25.04   &  21.12      &  22.08     &      1.59     &    -3.77   \\
   4   & 17:58:53.83   & -29:12:54.36   &  22.01      &  22.47     &     -6.22     &    -8.36   \\
   5   & 17:59:00.32   & -29:10:21.97   &  22.40      &  22.86     &      1.50     &    -1.99   \\
   6   & 17:59:00.78   & -29:12:06.03   &  23.32      &  23.77     &     -0.28     &    -4.08   \\
\enddata
\tablenotetext{a}{This table is available in its entirety in machine-readable
form in the online journal. A portion is shown here for guidance regarding its form and content.}
\end{deluxetable*}

\section{Summary}\label{sum}

We have identified for the first time a potential WD cooling sequence in the Galactic
bulge, based on  HST $F606W,F814W$-band images of the low-reddening Sagittarius
window, taken at two different epochs.  The WD sequence extends from $F606W \approx
22.5$ to 29 mag, with a color range of $0 \lesssim F606W - F814W \lesssim 1.5$ mag. 
Separating the disk and bulge stars through proper motions down to very faint
magnitudes was crucial in identifying the bulge WDs. We also identified a dozen
candidate EHBs, two ellipsoidal variables,  two dwarf novae and five candidate CVs.

The color spread of the WD sequence is rather large, and cannot be explained simply by
photometric errors, differential reddening and depth effects if all the WDs are CO-core. 
The large color spreads indicates that some of the observed WDs are be He-core, 
which are expected to be redder.

However, there are a few very red WDs ($F606W - F814W \gtrsim$ 1.0 mag), which cannot be
explained even by assuming they are very low-mass ($\sim 0.2 \,M_{\odot}$) He-core WDs.
Among the brighter of the very red objects we find one ellipsoidal variable, probably composed
of a WD accreting from a MS companion, and two dwarf novae. 
The fainter counterparts of these binaries could populate the region
where the reddest WDs  are observed in the CMD. These systems could be 
WDs which are in binary systems, composed of a
WD and a low-mass ($M < 0.3 \,M_{\odot}$) MS companion.  This hypothesis is further
supported by the discovery of five candidate CVs in the field,  based on the 2011--13
time-series photometry.  The dwarf novae and a fraction of bulge WDs also show a mild
$H_{\alpha}$-excess as expected.

A detailed analysis shows that the ratio of WD and MS star counts is about a factor of
two larger than  the ratio of CO-core WD cooling times and MS lifetimes if all the WDs
are CO-core. The observed number is consistent with the expected number if we assume
the presence of $\approx$ 30-40\% He-core WDs in the bulge, as already suspected
from the color spread of the WDs and the presence of CVs and ellipsoidal variables. 

\acknowledgments
This study was supported by NASA through grants GO-9750 and GO-12586 from the Space
Telescope Science Institute, which is operated by AURA, Inc., under NASA contract NAS~5-26555.
We would like to thank B. G{\"a}nsicke for useful suggestions and discussions about WDMS binaries. We would like to thank the anonymous referee for helpful suggestions
which led to an improved version of the paper.


\clearpage

\bibliographystyle{apj}
\bibliography{calamida}

\end{document}